# An Agent-Based Simulation of Residential Location Choice of Tenants in Tehran, Iran

**Ali Shirzadi Babakan and Abbas Alimohammadi**

**Abstract**

Residential location choice modeling is one of the substantial components of land use and transportation models. While numerous aggregated mathematical and statistical approaches have been developed to model the residence choice behavior of households, disaggregated approaches such as the agent-based modeling have shown interesting capabilities. In this paper, a novel agent-based approach is developed to simulate the residential location choice of tenants in Tehran, the capital of Iran. Tenants are considered as agents who select their desired residential alternatives according to their characteristics and preferences to various criteria such as the housing rent, accessibility to different services and facilities, environmental pollutions, and distance from their workplace and former residence. The choice set of agents is limited to their desired residential alternatives by applying a constrained NSGA-II algorithm. Then, agents compete with each other to select their final residence among their alternatives. Results of the proposed approach are validated by comparing simulated and actual residences of a sample of tenants. Results show that the proposed approach is able to accurately simulate the residence of 59.3% of tenants at the traffic analysis zones level.

## 1. Introduction

There are strong interactions between land use and transportation. For studying these interactions, numerous land use and transportation models have been developed. These models have been reviewed by many researchers (e.g. (Chang, 2006 ; Iacono et al., 2008)). One of the most important components of land use and transportation models is residential location choice modeling (Sener et al., 2011 ; Chang & Mackett, 2006). Many activities and urban travels of individuals are influenced by their residence. Thereby, residential location choice process of households can directly or indirectly affect various aspects of a city including transportation system, land uses, utilities, and socio-economic structures. Thus, this process has received high attention from many researchers in different fields such as urban planning, transportation, geography, and geosciences.

Residential location choice modeling has been initiated by Alonso (1960) and Lowry (1964) who applied economic and spatial interaction principles in their models. A decade later, Lerman (1976) and Mcfadden (1978) pioneered the use of discrete choice models in this area. In discrete choice models, still widely used in recent researches, utilities of a finite number of alternatives are calculated and the one with the maximum utility is selected. Within this class of models, multinomial logit (MNL) and nested logit (NL) are the most commonly used models (Sener et al., 2011 ; Rashidi et al., 2012). However, conventional models generally use zone-based aggregated characteristics of households and are insensitive to inherent heterogeneities among individual households. This issue has been recognized as one of the main sources of error in these models (Arentze et al., 2010 ; Benenson, 2004). Therefore, disaggregated models such as the microsimulation and agent-based models have found wider application.

In this paper, a novel agent-based approach has been developed to simulate the residential location choice of tenants in Tehran, Iran. Agent-based modeling have opened new ways to theoretically and experimentally model complex phenomena such as the urban system (Barros, 2004). An Agent-based model is composed of multiple interacting elements (agents) with some level of autonomy which can perceive their environment and act to change the environment according to their desires and objectives. These models almost have no limitations for directly representing and simulating behavior of urban elements including individuals and households (Pagliara & Wilson, 2010). They are "bottom-up" approaches, in which the behavior of system is emerged from the aggregation of agents' behavior. In agent-based residential location choice modeling, households can be represented as agents who decide to move and choose new dwellings and thereby affect behavior of other agents and urban components. In fact, from the perspective of agent-based modeling, regional or urban patterns of residential locations are outcomes of agents' residence choice behavior (Benenson, 2004).

There are many agent-based studies to model the housing market and residential location choice of households (e.g. Otter et al., 2001 ; Benenson, 2004 ; Jordan et al., 2012 ; Ettema, 2011 ; Rosenfield et al., 2013 ; Devisch et al., 2009 ; Haase et al., 2010 ; Gaube & Remesch, 2013 ; Jackson et al., 2008). In addition to these studies, a new generation of comprehensive urban models such as ALBATROSS[1] (Arentze & Timmermans, 2004), RAMBLAS[2] (Veldhuisen et al., 2000), MUSSA[3] (Martinez & Donoso, 2010 ; Martinez, 1996), ILUTE[4] (Salvini & Miller, 2005 ; Miller & Salvini, 2001), UrbanSim[5] (Waddell, 2002 ; Waddell et al., 2003), and ILUMASS[6] (Strauch et al., 2005) have been developed using microsimulation, cellular automata, and agent-based models. A detailed review of agent-based residential location choice models can be found in (Huang et al., 2014).

In almost all previous agent-based models, residential location choice is based on a utility-measuring or suitability-measuring function (Huang et al., 2014). There are rare studies (e.g. Jackson et al., 2008) which have utilized different heuristic or non-heuristic approaches in this area. Therefore, in this study, a novel two-step agent-based approach is proposed for residential location choice modeling which is the main contribution of this paper. In the first step of the proposed approach, a multi-objective decision making method, non-dominated sorting genetic algorithm II (NSGA-II), is introduced for evaluating residential alternatives by certain criteria and restricting the choice set of agents to a finite number of their desired residential alternatives. As far as the authors know, multi-objective decision making methods have not been used for determining a set of residential alternatives. For this purpose, two general approaches including consideration of all alternatives and random selection of some alternatives have been used by researchers. However, both approaches can raise some concerns. For example, the former unrealistically assumes that households search all alternatives, and the latter may result in inaccurate estimations (Rashidi et al., 2012). In addition to these general approaches, some researchers such as Rashidi et al. (2012) have proposed heuristic approaches to form smaller and more manageable choice sets. But, these approaches also have some limitations, for example, Rashidi et al. (2012) only considered average work distance for choice set formation, while there are several criteria such as property value and neighborhood characteristics that clearly affect the selection of residential alternatives. However, the proposed multi-objective decision making method in this paper allows modelers to use various criteria and objectives for determining a set of residential alternatives which leads to more realistic results.

In the second step, a heuristic competition mechanism is suggested in which agents compete with each other to select their final residence among their desired residential alternatives. This means that the residence choice of each agent is influenced by choices of the other agents. This is a critical issue which is not supported by the conventional models due to their inherent aggregated nature. As a result, in this paper, agents may not reside in their best residential option, because it may be previously occupied by another agent looking for a residence. This is more similar to the actual process of residential choice of households in the real world.

---

[1] A Learning-Based Transportation Oriented Simulation System
[2] Regional planning model based on the microsimulation of daily activity patterns
[3] A Land Use Model for Santiago City
[4] Integrated Land Use, Transportation, Environment
[5] Urban Simulation model
[6] Integrated Land-Use Modeling and Transportation System Simulation

The main objective of this paper is the spatially explicit simulation of residential location choice of individual tenants. The proposed approach is a useful tool for simulating the residential location choice behavior of individual households and spatially explicit distribution of different socio-economic categories of population. The approach then can be used by urban planners and policy makers to investigate effects of different plans and policies on these concerns.

As a case study, the proposed approach is implemented in Tehran, the capital of Iran. Although more than 50 years have passed since the beginning of residential location choice modeling in the world, there is little research conducted in this area in Tehran. Tehran contains a large number of tenants who change their residence every year, but the residential location choice behavior of this population has been never studied. Awareness of spatial distribution of different socio-economic categories of tenants and their residential choice behavior is a basic requirement for effective urban planning. In order to address this requirement, a practical microsimulation approach is proposed in this paper. The proposed approach can greatly help urban planners and policy makers in Tehran to simulate different land use and transportation scenarios and predict their impacts on residential location choice patterns of different socio-economic groups of population.

The rest of the paper is organized as follow. A background of rental residence choice in Tehran is presented in Section 2. Section 3 provides a detailed explanation of the proposed agent-based approach. In section 4, the proposed approach is implemented in Tehran metropolis and results are presented. Section 5 provides validation results of the proposed approach followed by the discussion and conclusion in section 6.

## 2. Background: Rental residence choice in Tehran, Iran

Tehran, the capital of Iran, with an area of about 750 $km^2$ is located at longitude of 51° 8′ to 51° 37′ and latitude of 35° 34′ to 35° 50′. According to the census conducted by the Statistical Center of Iran in 2011, Tehran's population is 7,803,883, composed of 2,245,601 households, of which about 950,000 are tenants (Tehran Municipality, 2013b). Tehran consists of 560 traffic analysis zones (TAZs) of which 532 zones include residential areas (Figure 1). In this paper, these zones are used as the spatial units for simulating residential location choice of tenants.

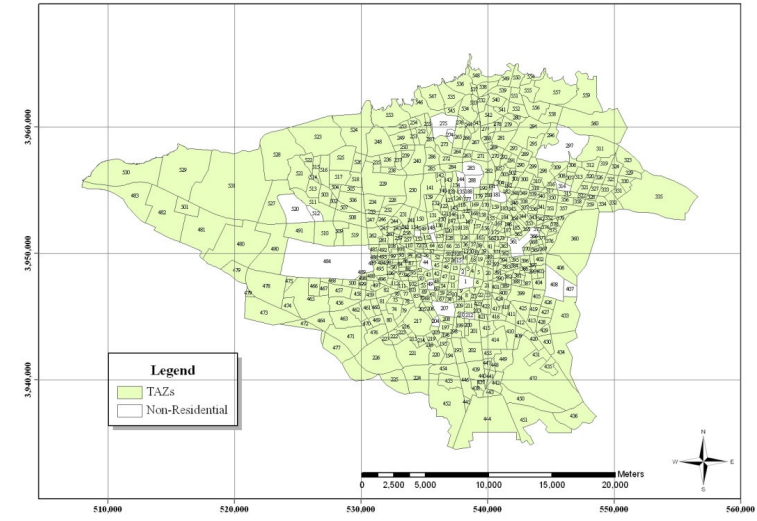

Figure (1): Spatial distribution of 560 traffic analysis zones (TAZs) of Tehran

The process of renting a residence in Iran is substantially different from those of the other countries. In Iran, landlords determine the rent of their properties by considering neighborhood characteristics and other attributes of the property. Then, usually the first tenant who affords and likes the property, rents it (Habibi & Ahari, 2005). Study of the process of renting a house in Tehran and consultation with real estate agencies show that landlords usually don't like to rent their properties to singles or households with many members. Therefore, if a number of households simultaneously ask to rent a property, couple households have a greater chance of success. The period of renting a house usually is one year in Iran. After one year, if the landlord and tenant don't reach an agreement for extension of the renting contract for the next year, the tenant has to leave his residence and to look for a new residence (Habibi & Ahari, 2005). According to the available statistics, about 30% of tenants in Tehran change their residence every year (Tehran Municipality, 2013b). This research is an attempt to simulate residential location choice process of these households. Therefore, in this paper, it has been assumed that agents want or have to change their residence and attempt to rent the best possible residence according to their characteristics and preferences.

It should be noted that due to some reasons such as having low incomes, low prices of fuel and public transit fares, high housing rents and existence of more employment opportunities in Tehran, some tenants prefer to reside in the surrounding cities of Tehran such as Karaj, Shahriar, Robat Karim, Eslamshahr, and Pardis, while their workplaces are located in Tehran. These cities are located in approximately short distances, less than 40 km, from Tehran. Therefore, these tenants commute between these cities and Tehran every day by their private car or public transit including bus, taxi and train. These tenants are not considered in this study. However, it is assumed that tenants who cannot reside in any residential zone of Tehran may have to move to one of these cities.

## 3. Proposed approach

In the proposed agent-based simulation of rental residence choice, tenants are simulated as agents who look for a preferred residence. They search among residential zones and select their appropriate residential alternatives by considering some criteria such as housing rent, environmental pollutions, distance from their workplaces and former residence, and accessibility to various services. Finally, to select their preferred residence, agents compete with each other. General framework of the proposed approach is shown in Figure (2). This framework consists of three basic modules including; I) generation of tenants (agents) using the Monte Carlo simulation, II) determination of desirable residential alternatives of agents using NSGA-II, and III) competition between agents to determine their final residence. These modules are briefly explained below.

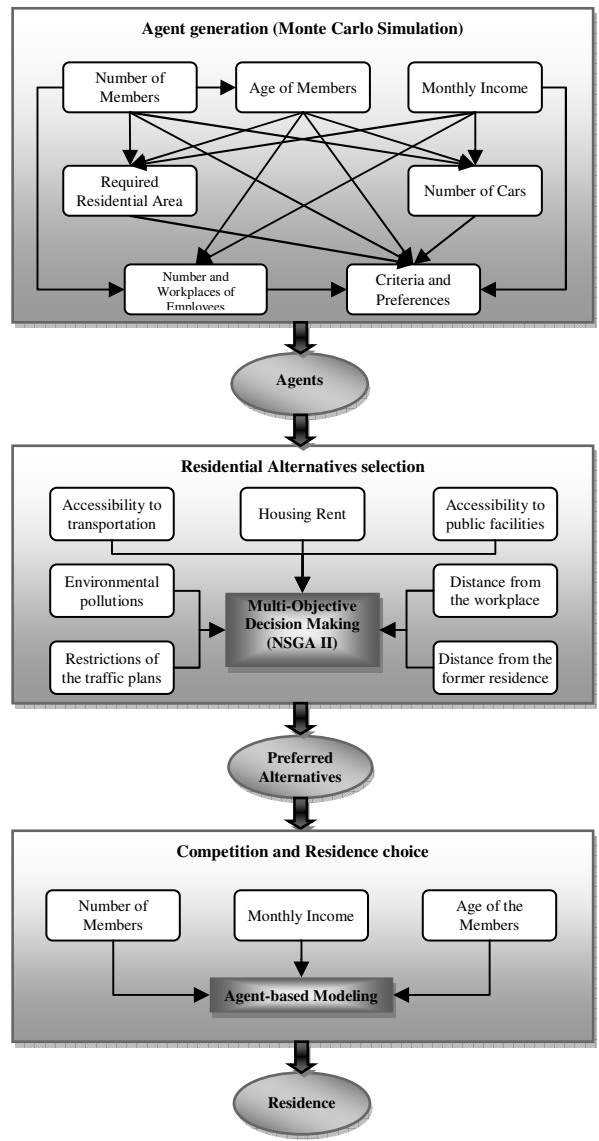

Figure (2): General framework of the proposed approach

## 3.1. Agent generation

The first module of the proposed approach is generation of tenants (agents) with required attributes for simulating their residential location choice behavior. Demographic and socio-economic attributes of agents are generated using the Monte Carlo simulation such that their aggregated average and standard deviation match with those of the available aggregated data in every zone (Tehran Municipality, 2013b). For this purpose, a sequential approach is used in which some attributes of households are simulated based on the previously determined attributes. This approach has been applied by Miller et al. (2004) in ILUTE model. In this approach, initially the number and age of members and monthly income of agents are generated using the Monte Carlo simulation such that the aggregated average and standard deviation of these attributes in each zone closely match to the available zone-based data of mean income, household size and percentages of different age groups provided by Tehran Municipality (2013b). Then, the number of cars and employees and the required residential area of agents are generated based on their previous attributes using the Monte Carlo simulation. Subsequently, using the available records of employment rate in different zones and general patterns of home-to-work travel distances (Tehran Municipality, 2013a), workplaces of employed members of agents are randomly allocated to zones. The employment capacity (EC) of each zone is computed by Eq. (1):

$$\text{Eq. (1):} \qquad EC = \frac{N_{ei}}{\sum_{i=1}^{n} N_{ei}} * N_a$$

where $N_{ei}$ is the total number of employments in zone $i$ and $N_a$ is the total number of agents.

Finally, residential criteria and preferences of agents are generated using the Monte Carlo simulation based on their previously simulated attributes. Depending on their demographic and socio-economic characteristics, agents use different criteria and preferences to select their residential zone. In order to simulate these criteria and preferences, stated preferences of 330 sample tenants with different characteristics were surveyed by filling questionnaires. Because of the lack of suitable sample data and limitations of collecting a large data set, a sample data consisting of 330 tenants were surveyed. Although this sample seems relatively small to represent the whole target population, due to surveying from different residential areas of Tehran, it covers tenants with various demographic, socio-economic and cultural characteristics. Statistical distribution of the sample data shows that it properly represents different categories of the target population (Table (1)). Sample tenants stated their preferences in three linguistic levels of importance including 'very important', 'important', and 'not important' in the questionnaires. It should be noted that tenants were quite justified to state 'important' and 'very important' preferences only for criteria which are actually considered by them in their residential location choice process. A summary of characteristics, preferred criteria, and percentage of tenants who stated 'very important' or 'important' preferences for each criterion is presented in Table (1).

Table (1): A summary of characteristics and preferred criteria of surveyed households

| Attribute | Category | Number | Percentage | Average Residential Area (m²) | Preferred Criteria (%) | | | | | | | | | | | | |
|---|---|---|---|---|---|---|---|---|---|---|---|---|---|---|---|---|---|
| | | | | | Housing Rent | Accessibility to Educational Locations | Accessibility to shopping Locations | Accessibility to Green and Recreational Locations | Accessibility to Cultural Locations | Accessibility to health Locations | Accessibility to Highways | Accessibility to Subway Stations | Accessibility to Bus Stations | Air and Noise Pollutions | Distance from the Workplace | Distance from the Former Residence | Without Traffic Restrictions |
| Size | Single | 28 | 8.5 | 55 | 100 | 17.9 | 14.3 | 7.1 | 10.7 | 3.6 | 71.4 | 35.7 | 17.9 | 32.1 | 92.9 | 46.4 | 46.4 |
| | Couple | 97 | 29.4 | 74 | 100 | 10.3 | 40.2 | 50.5 | 7.2 | 26.8 | 83.5 | 43.3 | 20.6 | 43.3 | 90.7 | 52.6 | 56.7 |
| | 3-4 | 171 | 51.8 | 86 | 100 | 71.3 | 33.9 | 57.3 | 8.8 | 34.5 | 87.1 | 52.0 | 22.8 | 41.5 | 84.2 | 62.6 | 53.8 |
| | > 4 | 34 | 10.3 | 97 | 100 | 88.2 | 29.4 | 64.7 | 5.9 | 44.1 | 85.3 | 52.9 | 20.6 | 41.2 | 79.4 | 61.8 | 52.9 |
| Monthly income (million IRR)* | < 7.5 | 69 | 20.9 | 67 | 100 | 42.0 | 18.8 | 34.8 | 5.8 | 21.7 | 65.2 | 53.6 | 29.0 | 20.3 | 92.8 | 68.1 | 39.1 |
| | 7.5-30 | 204 | 61.8 | 81 | 100 | 58.3 | 34.8 | 53.9 | 8.8 | 33.3 | 87.7 | 47.1 | 21.1 | 40.2 | 85.8 | 56.4 | 50.0 |
| | > 30 | 57 | 17.3 | 102 | 100 | 33.3 | 47.4 | 64.9 | 8.8 | 31.6 | 96.5 | 45.6 | 14.0 | 70.2 | 80.7 | 52.6 | 86.0 |
| Number of Cars | 0 | 34 | 10.3 | 71 | 100 | 55.9 | 38.2 | 44.1 | 8.8 | 26.5 | 26.5 | 91.2 | 73.5 | 26.5 | 97.1 | 55.9 | 2.9 |
| | 1 | 231 | 70.0 | 82 | 100 | 55.8 | 34.6 | 55.8 | 8.2 | 32.9 | 88.7 | 47.2 | 18.2 | 37.7 | 86.1 | 58.4 | 49.4 |
| | > 1 | 65 | 19.7 | 98 | 100 | 29.2 | 27.7 | 41.5 | 7.7 | 24.6 | 100 | 29.2 | 6.2 | 61.5 | 81.5 | 58.5 | 96.9 |
| Total | | 330 | 100 | 78 | 100 | 50.6 | 33.6 | 51.8 | 8.2 | 30.6 | 84.5 | 48.2 | 21.5 | 41.2 | 86.4 | 58.2 | 53.9 |

* IRR (Iranian Rial); At the time of this study 1USD is about 36000 IRR

### 3.2. Residential alternatives selection

In this step, agents freely select their desired residential alternatives using a multi-objective decision making method without considering the choice set of other agents and residential capacity of zones. They may have multiple and conflicting objectives. For example, they may want to reside in a zone with low rent and high accessibility to different services. But these objectives usually conflict with each other, because a high accessibility is generally coincided with a higher rent. Thus, each agent attempts to meet his objectives and accordingly selects the most suitable alternatives. In other words, each agent faces with a multi-objective decision making problem. In the proposed approach, a constrained NSGA-II is developed to determine a finite number of the best possible residential alternatives (up to ten alternatives in this case study) for agents in accordance to their residential criteria and preferences.

A number of different evolutionary algorithms such as MOGA[7] (Fonseca & Fleming, 1993), NSGA[8] (Srinivas & Deb, 1994), NPGA[9] (Horn et al., 1994), SPEA[10] (Zitzler & Thiele, 1998), PAES[11] (Knowles & Corne, 1999) and NSGA-II (Deb et al., 2002) have been developed to solve multi-objective optimization problems. Because of using elitism, SPEA, PAES and NSGA-II have attracted more interest of researchers. Zitzler et al. (2000) showed that elitism results in enhancing the convergence of a multi-objective evolutionary algorithm. NSGA-II is one of the fast and most efficient of elitist algorithms which has been widely used for solving multi-objective optimization problems in various applications (Li et al., 2010 ; Iniestra & Gutierrez, 2009 ; Huang et al., 2010). Deb et al. (2002) compared the convergence and spread of solutions of NSGA-II, PAES and SPEA on difficult test problems and found that NSGA-II is able to find better solutions for most problems.

---

[7] Multi-Objective Genetic Algorithm
[8] Non-dominated Sorting Genetic Algorithm
[9] Niched Pareto Genetic Algorithm
[10] Strength-Pareto Evolutionary Algorithm
[11] Pareto-Archived Evolution Strategy

In multi-objective optimization problems, there is not a global optimal solution with respect to all objectives. But there is a set of non-dominated solutions, generally known as the Pareto-optimal solutions. Since no one of the non-dominated solutions is better than the other, each of them can be accepted as optimal solution (Luh et al., 2003). NSGA-II uses the concept of non-domination to distinguish optimal solutions. For a multi-objective optimization problem as expressed in Eq. (2), it is said that solution *a* dominates solution *b* or *b* is dominated by *a* or *a* is not dominated by *b* if Eq. (3) is satisfied (Van Veldhuizen & Lamont, 2000):

$$Eq. (2): \quad minimize \quad f(x) = \{f_1(x), f_2(x), \ldots, f_n(x)\}$$
$$subject\ to \quad g(x) = \{g_1(x), g_2(x), \ldots, g_n(x)\}$$
$$Eq. (3): \quad if\ \forall i \in \{1,2,\ldots,n\}: f_i(a) \leq f_i(b) \quad and$$
$$\exists i \in \{1,2,\ldots,n\}: f_i(a) < f_i(b)$$

where:

*f(x)* is the set of objective functions,

*g(x)* is the set of constraints,

*a* and *b* are the possible solutions.

NSGA-II starts with an initial random generation of parent population, $P_0$ of size N. This population contains possible solutions of the multi-objective optimization problem. In this research, $P_0$ contains possible residential zones for each agent. Afterwards, an offspring population, $Q_0$ of size N, is generated by applying genetic operators including binary tournament selection, crossover and mutation on $P_0$. Since the procedure is repetitive after the initial generation, the *t*th generation is described at the following. A combined population, $R_t$ of size 2N, is generated by $R_t=P_t \cup Q_t$. The solutions in $R_t$ are ranked and assigned to different fronts according to the non-dominated sorting. In this sorting process, solutions which do not dominate each other and dominate all the other solutions are assigned to the first (best) front. This process is continued until all solutions of $R_t$ are assigned to the non-dominated fronts. Then, the parent population in the next generation, $P_{t+1}$ of size N, is generated from the solutions belonging to the first (best) fronts of $R_t$. The new offspring population, $Q_{t+1}$ of size N, is generated by applying binary tournament selection, crossover and mutation operators on $P_{t+1}$. In the binary tournament selection, the winner (better) solution is selected using the crowded-comparison operator. In this operator, a solution with lower rank is selected as the winner. If two solutions have the same rank, a solution with higher crowding distance is declared as the winner. The crowding distance of a solution is the perimeter of the cuboid formed by its nearest neighboring solutions in the objective space. In fact, this distance estimates the density of solutions surrounding a particular solution. The algorithm is continued until the convergence criterion such as the maximum number of generations is satisfied and $P_t$ is returned as the output (Deb et al., 2002). $P_t$ contains the optimal residential alternative zones of each agent. More details of this algorithm and its implementation procedure can be found in Deb et al. (2002). .

In the following of this section, the most important criteria for residential location choice of tenants derived from the survey of stated preferences of sample tenants in Tehran (Table 1) are described in detail. It should be noted that depending on their preferences, agents may use one, some, or all of these criteria for selecting their desired residential alternatives.

### 3.2.1. Housing rent

Housing price is one of the most important factors affecting the residential location choice of households (Hunt, 2010 ; Ettema, 2011 ; Devisch et al., 2009 ; Jackson et al., 2008 ; Wu et al., 2013 ; Lee & Waddell, 2010 ; Waddell et al., 2003 ; Sener et al., 2011). In this research, it is assumed that the average housing rent per square meter in every zone is exogenously known and fixed during the simulation. Tenants only select zones in which the housing rent is compatible with their affordability and required residential area. For this purpose, a condition is considered to limit the search space of agents to zones in which the housing rent of their required residential area is between the specified minimum and maximum percentages of their monthly income. Because of limitations caused by the income level of agents, the maximum limit seems applicable for all agents. But, the minimum limit is considered for some agents, especially for agents with high monthly incomes, because, for cultural and social reasons, they usually prefer to reside in rich neighborhoods (Habibi & Ahari, 2005). As a result, these limits lead to consideration of socio-economic composition of residents in the residential location choice process of agents. Also, by applying this condition, the search space of agents and required computational time are significantly decreased. The following objective function is considered for each agent:

Eq. (4): $$\min_{\substack{\forall a \in N_a \\ \forall i \in N_z}} (RA_a * R_i) \quad where: Pmin_a * I_a \leq RA_a * R_i \leq Pmax_a * I_a$$

where:

$RA_a$ is the required residential area of agent $a$;

$R_i$ is the average housing rent per square meter in zone $i$;

$I_a$ is the agent's monthly income;

$Pmin_a$ and $Pmax_a$ respectively are the minimum and maximum percentages of monthly income which are considered by agent $a$ for renting a residence. These percentages are defined using the Monte Carlo simulation;

$N_a$ is the set of agents that the criterion is important for them; and

$N_z$ is the set of residential zones.

### 3.2.2. Accessibility to public facilities

Households generally prefer to live in zones with high accessibilities to public facilities including educational (Hunt, 2010 ; Jackson et al., 2008 ; Myers & Gearin, 2001 ; Wu et al., 2013 ; Lee & Waddell, 2010), shopping (Hunt, 2010 ; Lee & Waddell, 2010 ; Chen et al., 2008 ; Lee et al., 2010 ; Srour et al., 2002 ; Sener et al., 2011), green and recreational (Wu et al., 2013 ; Chen et al., 2008 ; Srour et al., 2002 ; Sener et al., 2011), cultural (Sener et al., 2011), and health locations (Wu et al., 2013). However their preferences for accessibility to various facilities are different depending on their socio-economic and demographic characteristics. For example, while accessibility to schools may be very important for a tenant with student members, it may not be important for the others. Overall accessibility of zones to public facilities for each agent who this criterion is important for him is calculated by Eq. (5) which is an extension of the accessibility index developed by Tsou et al. (2005). Also, accessibility of zones to each public facility type is calculated using this equation, where value of $p_{ka}$ for the public facility type of interest is set to 1 and values of $p_{ka}$ for the other public facility types are set to 0. This index is normalized using $(x-x_{min})/(x_{max}-x_{min})$.

$$\text{Eq. (5):} \quad \max_{\substack{\forall a \in N_a \\ \forall i \in N_z}} \left( \sum_k \sum_{j(k)} p_{ka} * w_{j(k)} * d_{ij}^{-2} \right)$$

where:

$k$ is the type of public facility including educational, shopping, green and recreational, cultural, and health locations;

$j(k)$ is the $j$th case of the $k$th type of public facility;

$p_{ka}$ is the agent's preference to the public facility type $k$ which is determined in the agent generation module, where $\sum p_{ka}=1$;

$w_{j(k)}$ is the relative effect of $j(k)$ which is calculated by $w_{j(k)}=A_{j(k)}/max(A_{j(k)})$, where $A_{j(k)}$ is the area of $j(k)$;

$d_{ij}$ is the distance between zone $i$ and $j(k)$.

### 3.2.3. Accessibility to transportation services

Like the accessibility to public facilities, households usually prefer to reside in zones with high accessibilities to transportation services including highways and/or public transit (Hunt, 2010 ; Myers & Gearin, 2001 ; Wu et al., 2013 ; Sener et al., 2011). Overall accessibility of zones to transportation services for each agent who this criterion is important for him is measured by Eq. (6) which is derived from the studies of Currie (2010) and Delbosc and Currie (2011). Also, this equation is used to calculate accessibility of zones to each transportation service type, where value of $P_{ka}$ for the transportation service type of interest is set to 1 and values of $P_{ka}$ for the other transportation service types are set to 0. Finally, accessibilities are normalized using $(x-x_{min})/(x_{max}-x_{min})$.

$$\text{Eq. (6):} \quad \max_{\substack{\forall a \in N_a \\ \forall i \in N_z}} \left( \sum_k \sum_{t(k)} \frac{A_{t(k)}}{A_i} * P_{ka} \right)$$

where:

*k* is the type of transportation service including highways, bus stops and subway stations;

*t(k)* is the *t*th case of the *k*th type of transportation service;

$A_{t(k)}$ is the area of service range of *t(k)* which is inside zone *i*;

$A_i$ is the area of zone *i*; and

$P_{ka}$ is the agent's preference to the transportation service type *k* which is determined in the agent generation module, where $\sum P_{ka}=1$;

A fundamental component of accessibility to transportation services is the service range or access distance. Researchers have typically used various walking distances ranging from 300 to 800 m for this distance. Validity of these heuristic distances has been investigated using the travel survey data (Mavoa et al., 2012). However, these distances are longer in Tehran, because in addition to walking, people often use local taxis with low fares to access transit services. Therefore, according to the comprehensive transportation and traffic studies of Tehran (Tehran Municipality, 2013a), the access distances to bus stops, subway stations, and highways are defined as 1.5 km, 1.9 km, and 2 km, respectively.

### 3.2.4. Air and noise pollutions

Air and noise pollutions can have important influences on the residential location choice of households (Hunt, 2010). By using the annual mean noise and air pollution records of Tehran (Tehran Municipality, 2013a), residential zones are classified in five categories varying from clean to highly polluted zones. Then, residential alternatives of agents having very important preferences for air and noise pollutions are restricted to medium to clean pollution classes. Also, the following objective functions are applied for agents who these criteria are recognized as their important criteria.

$$\text{Eq. (7):} \quad \min_{\substack{\forall a \in N_a \\ \forall i \in N_z}} (AP_i)$$

$$\text{Eq. (8):} \quad \min_{\substack{\forall a \in N_a \\ \forall i \in N_z}} (NP_i)$$

where $AP_i$ and $NP_i$ respectively are the annual mean air and noise pollutions of zone *i*.

### 3.2.5. Distance from the workplace

Distance of residence from the workplace is an important factor for many households (Hunt, 2010 ; Jackson et al., 2008 ; Myers & Gearin, 2001 ; Wu et al., 2013 ; Rashidi et al., 2012 ; Lee & Waddell, 2010 ; Chen et al., 2008 ; Lee et al., 2010 ; Srour et al., 2002 ; Waddell et al., 2003). The following objective function is used for agents who prefer to minimize distance of their residence from the workplace(s) of their members.

$$\text{Eq. (9):} \quad \min_{\substack{\forall a \in N_a \\ \forall i,w \in N_z}} \sum_{m=1}^{n} d_{iw(m)}$$

where:

$d_{iw(m)}$ is the distance between zone $i$ and workplace of employed member $m$ of agent $a$.

$n$ is the number of employees in agent $a$.

### 3.2.6. Distance from the former residence

Another important factor in residential location choice of some households is distance from their former residence (Jackson et al., 2008 ; Chen et al., 2008). Because of familiarity, various dependencies, and meeting their requirements in the former residential area, some agents prefer to live nearby their former residential area. Objective function of these agents is defined as:

$$\text{Eq. (10):} \quad \min_{\substack{\forall a \in N_a \\ \forall i,Fr \in N_z}} (d_{iFr(a)})$$

where $d_{iFr(a)}$ is the distance between zone $i$ and the former residence of agent $a$.

### 3.2.7. Traffic restrictions

Households usually consider type of streets and traffic situations in their residential area (Hunt, 2010 ; Martinez & Viegas, 2008). Because of traffic restrictions such as the odd-even car restrictions and restrictions on private cars in central areas of Tehran, some households prefer to reside outside these areas. For this purpose, agents with very important preferences to traffic restrictions look for their residence in non-restricted areas. Also, the following objective function is used to minimize selection of traffic restricted zones by agents having important preferences to these restrictions:

$$\text{Eq. (11):} \quad \min_{\substack{\forall a \in N_a \\ \forall i \in N_z}} (TR_i)$$

where $TR_i$ is the traffic restriction code of zone $i$ with the values of 0 for no traffic restriction, 1 for odd-even car restriction, and 2 for restriction on all private cars.

### 3.3. Competition and residence choice

After selection of desired residential alternatives, agents compete with each other to select their final residence. In reality, tenants relocate in different months of the year. A monthly time step is considered in this paper, because rent is usually paid each month and tenants commonly begin to look for a new residence from one month before their lease deadline. In other words, households usually find a new residence during one month. For simulation of this process, the available statistical information of the residential relocation rates in different months of the year in Tehran (Tehran Municipality, 2013a) is used to randomly define the relocation month of each agent. At the other hand, residential capacity of zones is limited in each month and estimated by Eq. (12):

$$\text{Eq. (12):} \qquad RC_{it} = \frac{RA_i}{\sum_i RA_i} * (\alpha_t * N_{at})$$

where:

$RC_{it}$ is the residential capacity of zone *i* at month *t*;

$RA_i$ is the total residential area inside zone *i*;

$N_{at}$ is the number of agents who look for residence at month *t*; and

$\alpha_t$ is a balancing parameter that presents an equilibrium between the residential supply and demand at month *t*. For example, value of more than one for this parameter means that the residential supply is more than the demand. This parameter is calibrated using the available statistical information in each month (Tehran Municipality, 2013a).

In each month, a number of agents compete with each other for selecting a residence among their residential alternatives. They search among their desired alternative zones according to the distance of these zones from their former residence and reside in the first zone having the enough residential capacity. If demand for a zone is more than its capacity, then agents compete with each other. As mentioned in section 2, study of the process of renting a house and consultation with real estate agencies in Tehran show that landlords commonly prefer to rent their properties to households with few members (except singles), without child, and with high incomes. Therefore, in the proposed residential competition, agents with fewer members (except singles), with higher incomes, and without child have a higher chance of success, respectively. In the case of equity of the above mentioned conditions, winners are randomly selected. Defeated agents try to reside in their next alternatives and the process is repeated until all agents either reside in one of their residential alternatives or evaluation of all alternatives of defeated agents is completed. Agents who cannot reside in any of their alternatives are moved to the next month. It is assumed that these agents can take additional time (up to one month) from the landlords to find a new residence. Landlords usually agree with this request, because the legal process of expelling a tenant from occupying a property generally is too long and expensive. Therefore, these agents would compete again with agents of the next month and if they still cannot reside in any zone, no residence is allocated for them. In fact, they are agents with many members or low incomes who cannot reside in any zone of Tehran and possibly they have to move to the surrounding cities of Tehran.

## 4. Empirical results

In an attempt to cover all socio-economic categories of tenants in Tehran, 100,000 agents were simulated using the agent generation module developed in MATLAB 7.9.0 software. These agents were randomly classified to 12 months according to the percentage of residential relocation in different month of the year in Tehran (Tehran Municipality, 2013a). Table (2) shows the number of agents who look for a residence in each month. As shown in this table, due to school holidays, about 65% of the annual residential relocation in Tehran occur in three months of the summer (Tehran Municipality, 2013a).

Table (2): Distribution of residential relocation and agents in different months of the year

| Time Period (Month) | Percentage Of Relocation | Number Of Agents |
|---|---|---|
| April | 1% | 1000 |
| May | 8% | 8000 |
| June | 7% | 7000 |
| July | 19% | 19000 |
| August | 22% | 22000 |
| September | 24% | 24000 |
| October | 8% | 8000 |
| November | 5% | 5000 |
| December | 2% | 2000 |
| January | 1% | 1000 |
| February | 2% | 2000 |
| March | 1% | 1000 |
| **1 Year** | **100%** | **100000** |

Agents then selected up to 10 residential alternatives by the developed NSGA-II. For determining the maximum number of residential alternatives of agents, the proposed approach was run with different maximum numbers of alternatives varying from 2 to 15 for sample tenants presented in Table (1). Simulation accuracy of the actual residential zone of these tenants by different maximum numbers of alternatives has been shown in Figure (3). As seen in this Figure, the simulation run with the maximum number of ten residential alternatives shows the best performance. Also, because of using relatively large zones in this case study, it seems incredible that agents be able to search more than ten zones.

In order to test the repeatability and stability, the proposed simulation approach was executed 5 times and aggregated results of these runs were compared with each other. Results showed that aggregated residential location choice behaviors of different demographic and socio-economic categories of agents in different simulation runs are very close to each other with differences of less than 2%. This suggests acceptable repeatability and stability of the proposed approach. Therefore, in this paper, empirical results of one simulation run are presented. The average run time of each simulation is about 7.3 hours in a platform with Core i7 2.00 GHz of CPU and 6 GB of RAM. The simulation stops either all agents reside in one of their residential alternatives or evaluation of all alternatives of agents is completed.

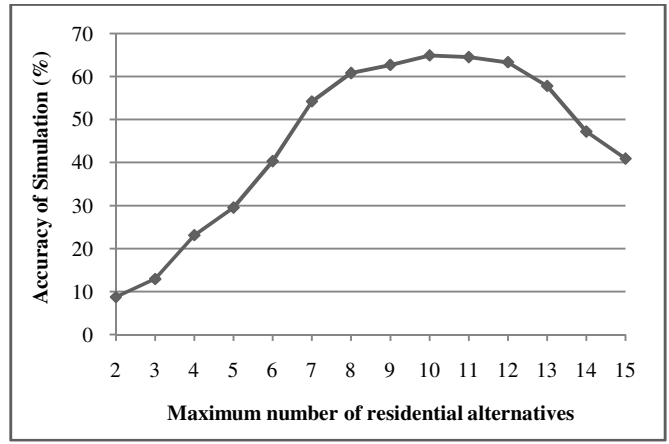

Figure (3): Accuracy of simulation of the actual residential zone of 330 sample tenants by the maximum number of residential alternatives selected by them

In the following of this section, major patterns derived from simulation of the residential location choice of individual agents are analyzed. For different demographic and socio-economic categories of agents, these aggregated patterns are described by various criteria considered in the simulation design including housing rent, accessibility to transportation services, accessibility to public facilities, distance from the former residence and workplace, air and noise pollutions and traffic restrictions. In fact, this analysis shows that how well the proposed approach can simulate the observed or expected residential location choice behaviors of tenants in Tehran. This highlights the performance of the simulation in generating expected and near-reality results.

Spatial distribution of residential alternatives of agents is shown in Figure 4(a). As expected, because of low housing rents and high accessibilities to employment opportunities, public facilities and public transit services, central areas of Tehran are more attractive than the other areas. However, relatively higher air and noise pollutions and traffic restrictions in these areas lead to reducing the residential attractiveness of these areas for some agents, especially for agents with high incomes that is consistent with the observations. Also, spatial distribution of final residence of agents is shown in Figure 4(b). As illustrated in this Figure, the population density is higher in central areas which is consistent with the existing population distribution in Tehran (Tehran Municipality, 2013b). Due to higher residential demand and consequently more intense competition with other agents in central areas, 64.2% of residents of these areas cannot reside in their first three alternatives. On the other hand, because of low residential demands in southern and northwestern areas, 69.5% of residents of these areas have resided in one of their first three alternatives.

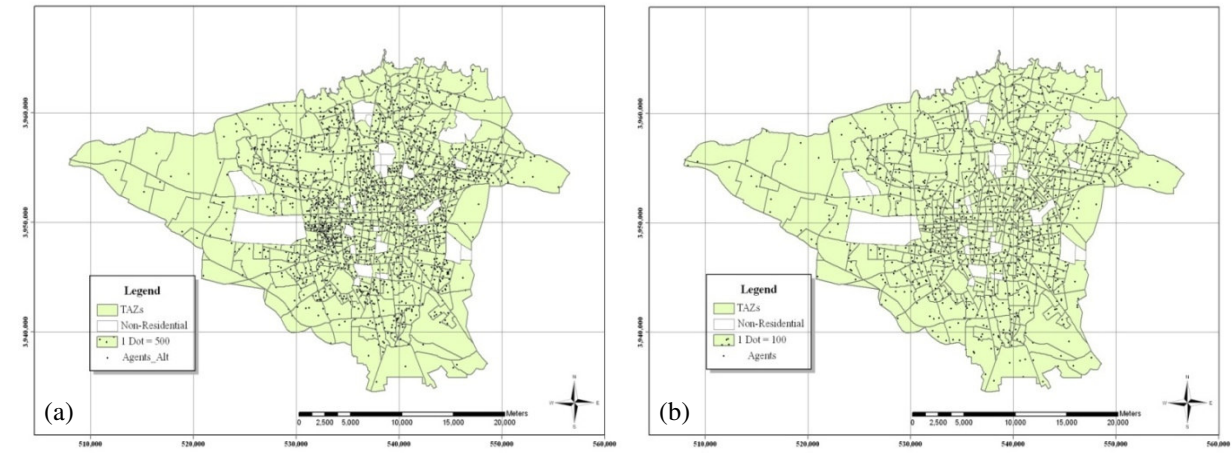

Figure (4): Distribution of: (a) the residential alternatives; and (b) the final residence of agents in TAZs

Distribution of agents by the number of their residential alternatives has been represented in Figure 5(a). As can be seen, 83.3% of agents have selected the maximum possible number of residential alternatives, i.e. ten alternatives. This suggests that the constraints applied in the proposed approach are reasonable so that the choice set of agents is limited to a few residential alternatives. However, the choice set of 1.8% of agents has been limited to less than 4 residential alternatives. These agents generally have low incomes or strict preferences for choosing their residential location. Also, distribution of agents by the rank of their final residence among their residential alternatives has been represented in Figure 5(b). As indicated in this histogram, 19.7% of agents have resided in their first residential alternative. These agents generally have a few members and high incomes or have selected their residence from southern and northwestern areas in which the residential demand is less than those of the other areas. An observable decreasing trend in this histogram is consistent with the expectations. Because, agents attempt to reside in their first alternatives and the number of agents who have resided in their last alternatives has sequentially decreased. Finally, only 0.7% of agents cannot reside in any zone and may have to move to the surrounding cities of Tehran. It seems that low income level and high household size of these agents are the main reasons for this situation. Mean income and household size of these agents are 7,020,000 IRR[12] and 4.3, respectively.

Distribution of agents by distance of their residence from their workplace(s) is shown in Figure 6(a). Although proximity to the workplace is important for 86.4% of agents, results of the simulation show that 21.2% of agents have resided in distances of more than 10 kilometers from their workplace(s). This unexpected result suggests that though a great number of agents prefer to reside in proximities of their workplace(s), a majority of them cannot practically reside close to their workplace(s) due to other important factors such as their socio-economic characteristics, housing rent, and not meeting the other requirements. In addition, low prices of fuel and transit fares in Tehran may be regarded as the other important reason for residing of a considerable number of agents far from their workplace(s). However, because of reduction of their commuting costs, low-income agents and who have two or more employees with near workplaces have shown higher interests to reside close to their workplace(s) which is consistent with the expectations. As an example, 65.7% of agents who have two or more employees with near workplaces have resided in distances of less than 5 km from their workplaces, whereas only 8.1% of these agents have resided in distances of more than 10 km from their workplaces.

---

[12] Iranian Rial. At the time of this study 1 USD is about 36,000 IRR.

Figure 6(b) shows distribution of agents by distance between their current and former residences. As expected, due to some reasons such as forced relocation because of lack of agreement with the landlord for extension of the lease, familiarity and dependency to the former residential area, and high financial and psychological costs of moving to far areas, a great number of agents have preferred to reside in proximities of their former residence. For instance, only 4.7% of agents have moved to distances of more than 10 km from their former residence.

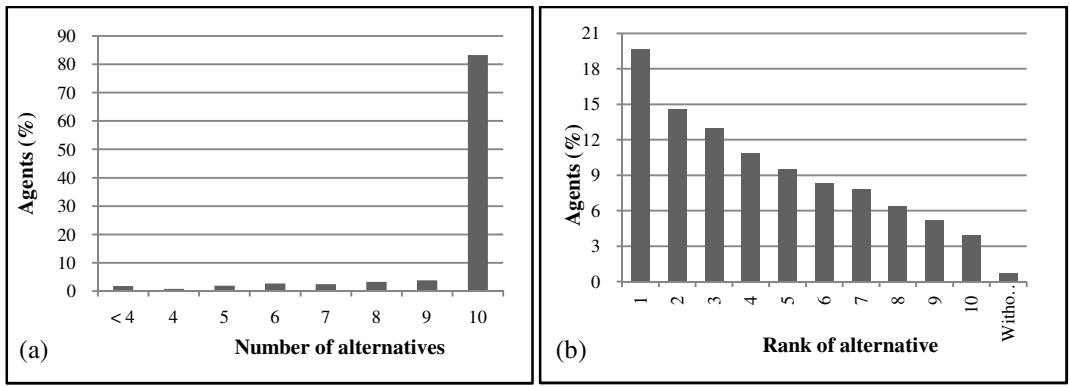

Figure (5): Distribution of agents by: (a) the number of their residential alternatives; (b) the rank of their final residence among their alternatives

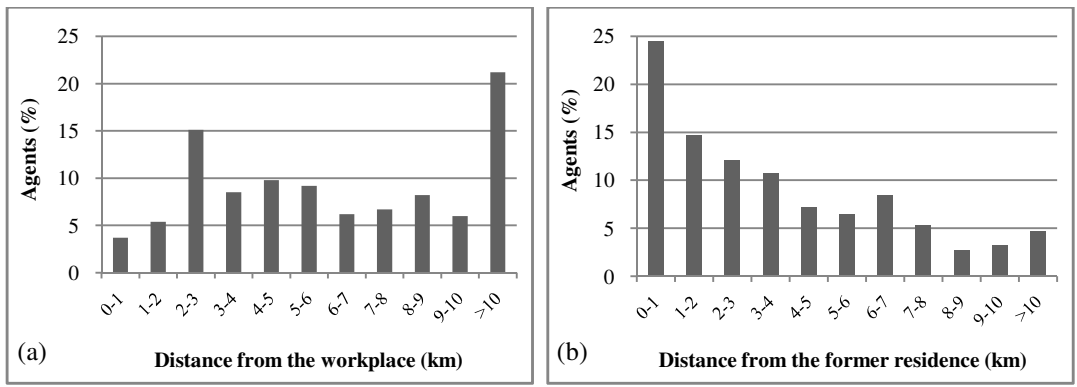

Figure (6): Distribution of agents by: (a) distance from their workplace; (b) distance from their former residence

Spatial distribution of the mean income of agents exhibits a clear south-north pattern (Figure (7)). This pattern shows considerable similarities with the existing distribution of mean income of residents in Tehran (Tehran Municipality, 2013b). As hypothesized, agents have resided in zones in which the mean income of residents is compatible with their income level. For example, although high-income agents could easily reside in zones with low housing rents, they have preferred to reside in zones with higher housing rents compatible with their income level.

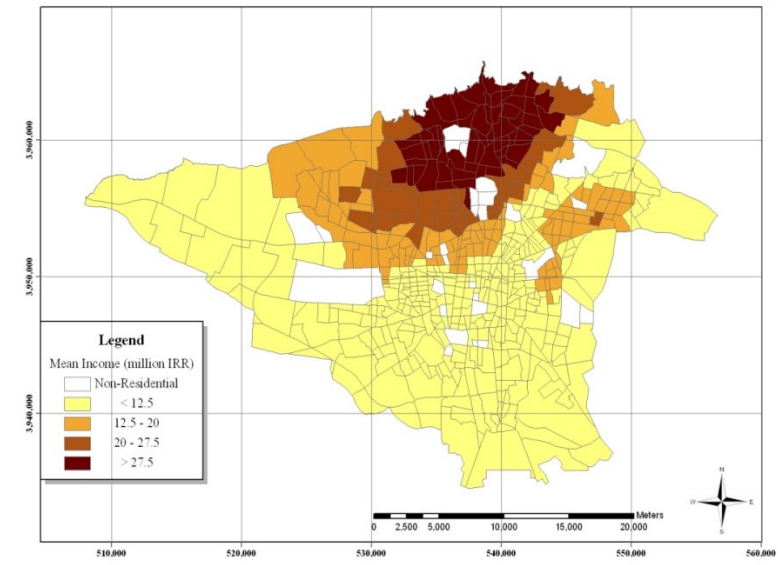

Figure (7): Distribution of the mean income of residing agents in TAZs

As expected, heterogeneous distributions of highways and public transit in Tehran lead to high importance of these factors in residential location choice of agents. Simulation results show that agents with cars particularly with more than one car have more resided in proximities of highways. For example, 71.8% of agents having more than one car have resided in zones with accessibilities of more than 70% to highways. On the other hand, as anticipated, agents without car have considerably resided in central zones with higher accessibilities to public transit services. In addition, an unanticipated weak association is observed between residing in zones with high accessibilities to public transit and the number of members and employees of agents. Probably, because of higher demand for public transit use by their members, agents with more than one employee or three members slightly have more resided in proximities of public transit services.

As hypothesized, results show the importance of accessibility to various public facilities in residence choice of different socio-economic categories of agents. A majority of agents have selected their residential alternatives from zones with high accessibilities to shopping and recreational locations which is consistent with the expectations. For example, the number of residential alternatives selected by agents in zones with accessibilities of higher than 70% to shopping and recreational locations is 10.2% and 5.9% more than the average number of residential alternatives in all zones, respectively. In addition, as anticipated, agents with student members have tended to reside in proximities of educational locations. For example, 54.1% of agents having more than one student have resided in zones with accessibilities of higher than 70% to educational locations. Finally, results do not show any meaningful relationships between residential location choice and accessibility to cultural and health centers. This suggests the minor role of these criteria in rental residence choice in Tehran.

As expected, air pollution and especially noise pollution play an important role in the residence choice of agents.. An anticipated strong relationship between the income level of agents and their preferences for less polluted areas is observed so that high-income agents have mostly resided in zones with low air and noise pollutions. For example, only 13.8% and 7.7% of high-income agents have resided in zones with high levels of air and noise pollutions, respectively. On the other hand, low-income agents have been forced to reside in zones with higher pollutions due to their limited affordability and lower housing rents in these zones. Importance of the noise pollution can be clearly shown in residential areas around the Mehrabad airport. Although these areas have high accessibilities to highways, public transit services and public facilities, the number of residential alternatives selected by agents in these areas is 44.3% less than the average number of residential alternatives in all zones. Also, only 8.7% of residents of these areas have high income levels.

As hypothesized, results show the significant influences of traffic restrictions (e.g. restrictions for all and odd-even cars) on the residence choice of agents, especially on car-owning agents. The number of residents having one and more than one car in traffic-restricted areas respectively is 10.6% and 52.2% less than the average number of this class of residents in all zones. On the other hand, due to central location of these areas and high accessibilities to employment opportunities and public transit services, agents without car have shown great interests for residing in these areas which is consistent with the observations. Number of residents without car in these areas is 57.8% more than the average number of this class of residents in all zones. Finally, it should be noted that simulated distribution of the mean car ownership is considerably consistent with the existing mean car ownership distribution in Tehran (Tehran Municipality, 2013b).

In general, it can be said that major residential location choice patterns of different demographic and socio-economic categories of agents are considerably consistent with the observations and expectations. A summary of residential location choice and selected residential zones by agents have been presented in Tables (3) and (4).

Table (3): A summary of residential location choice of agents

| Category of Agents | Percentage of Agents (%) | Average Income (million IRR) | Average Number of Members | Average Car Ownership |
|---|---|---|---|---|
| Selected all 10 alternatives | 83.3 | 14.2 | 3.43 | 1.02 |
| Selected more than 7 alternatives | 92.8 | 11.4 | 3.49 | 0.97 |
| Selected less than 4 alternatives | 1.8 | 7.0 | 3.86 | 0.82 |
| Resided in the first alternative | 19.7 | 15.9 | 3.02 | 1.14 |
| Resided in one of the first three alternatives | 47.3 | 13.7 | 3.31 | 1.00 |
| Resided in one of the last three alternatives | 15.5 | 10.3 | 3.78 | 0.88 |
| Moved to the next time period | 5.7 | 7.1 | 4.05 | 0.84 |
| unable to reside in any zone | 0.7 | 6.7 | 4.24 | 0.76 |
| Resided in a zone with distance of less than 5 km from the former residence | 69.2 | 12.6 | 3.48 | 0.97 |
| Resided in a zone with distance of more than 10 km from the former residence | 4.7 | 13.5 | 3.46 | 1.02 |
| Resided in a zone with distance of less than 5 km from the workplace | 42.5 | 14.1 | 3.47 | 0.90 |
| Resided in a zone with distance of more than 10 km from the workplace | 21.2 | 14.9 | 3.50 | 1.05 |

Table (4): A summary of selected residential zones by agents

| Category of Residential Zones | Zone Number | Number of Agents | *Housing Rent per m² (%) | Air Pollution | Noise Pollution | **Overall Access to Public Facilities (%) | ***Overall Access to Highway Network (%) | ****Overall Access to Public Transit (%) |
|---|---|---|---|---|---|---|---|---|
| The most selected zones as residential alternatives by agents | 531 | 5639 | 62.7 | Clean | Clean | 56.4 | 70.6 | 37.3 |
|  | 311 | 5303 | 58.5 | Relatively Clean | Clean | 60.1 | 63.8 | 46.5 |
|  | 528 | 5058 | 68.3 | Clean | Clean | 43.2 | 65.2 | 29.6 |
| The least selected zones as residential alternatives by agents | 2 | 764 | 40.6 | Highly Polluted | Highly Polluted | 68.3 | 28.4 | 78.9 |
|  | 16 | 809 | 44.2 | Highly Polluted | Highly Polluted | 62.8 | 33.5 | 81.3 |
|  | 17 | 855 | 43.8 | Highly Polluted | Highly Polluted | 65.0 | 32.1 | 79.5 |
| Zones with the most residents | 531 | 828 | 62.7 | Clean | Clean | 56.4 | 70.6 | 37.3 |
|  | 560 | 814 | 67.4 | Clean | Clean | 58.9 | 66.4 | 41.2 |
|  | 360 | 790 | 38.6 | Medium | Relatively Clean | 41.2 | 58.0 | 33.4 |
| Zones with the lowest residents | 2 | 64 | 40.6 | Highly Polluted | Highly Polluted | 68.3 | 28.4 | 78.9 |
|  | 14 | 71 | 39.7 | Highly Polluted | Highly Polluted | 56.1 | 25.6 | 84.4 |
|  | 15 | 74 | 41.3 | Highly Polluted | Highly Polluted | 59.0 | 30.7 | 79.7 |
| Zones which were occupied earlier than other zones | 311 | 595 | 58.5 | Relatively Clean | Clean | 65.1 | 63.8 | 46.5 |
|  | 508 | 237 | 60.9 | Relatively Clean | Medium | 73.3 | 68.9 | 61.5 |
|  | 221 | 280 | 30.2 | Medium | Medium | 58.4 | 43.5 | 48.7 |

\* The housing rent per m² has been normalized using $(x-x_{min})/(x_{max}-x_{min})$.
\*\* Overall access to public facilities is calculated using Eq. (5) where all public facilities have the same preferences ($p_{ka}$). This index has been normalized using $(x-x_{min})/(x_{max}-x_{min})$.
\*\*\* Overall access to highway network is calculated using Eq. (6) where $P_{ka}$ is equal to 1 and k is highway. This index has been normalized using $(x-x_{min})/(x_{max}-x_{min})$.
\*\*\*\* Overall access to public transit is calculated using Eq. (6) where $P_{ka}$ is equal to 0.5 and k is the bus and subway. This index has been normalized using $(x-x_{min})/(x_{max}-x_{min})$.

## 5. Validation

One of the greatest challenges of utilizing agent-based models is their validation (Crooks & Heppenstall, 2012). In this research, 1350 tenant households with various socio-economic characteristics residing in different zones of Tehran were sampled for validation of the proposed approach. Socio-economic characteristics, residence and workplace(s) of these households were collected, but their residential criteria and preferences were simulated using the proposed approach. For this purpose, households were classified to 12 months according to the month in which they rented their residence. Then, residence of these households was simulated by the proposed approach and results were compared with those of the actual residence.

As shown in Figure (8), spatial distribution of simulated and actual residences of these households is significantly compatible. Distribution of households by distance of their simulated residence from their actual residence demonstrates validity of the simulation (Figure 9(a)). As indicated in this diagram, simulated residence of 60.2% of households is located in distance of less than one kilometer from their actual residence. Also, simulated residence of only 5.4% of households is located in more than ten kilometers from their actual residence. In addition, distribution of the actual and simulated residences of households by housing rent, proximity to their workplace(s) and distance from their former residence show high similarities (Figures 9(b), 10(a) and 10(b)).

Comparison of the actual and simulated residences of households in different spatial and attribute categories has been presented in Table (5). As shown in this table, the proposed approach has exactly simulated the residential zone of 59.3% of households. Also, simulated residential zone of 67.5% of households has been located in the adjacency of their actual residential zone, composed of the actual residential zone and its neighboring zones. In addition, the actual residence of 72.8% of households has been simulated as one of their residential alternatives. The approach is also able to simulate different attributes of the actual residence of more than 70% of households including the housing rent, air and noise pollutions and accessibilities to public facilities and transportation services with errors of less than 15%. For example, the housing rent of 84.0% of households has been simulated with errors of less than 15%. Therefore, it seems that the proposed approach shows promising performance.

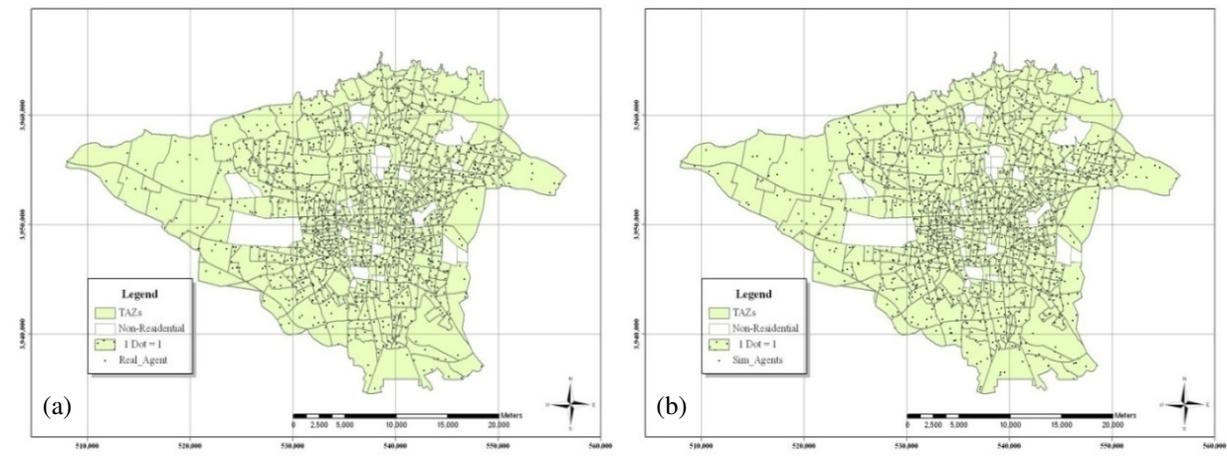

Figure (8): Distribution of (a) the actual; and (b) simulated residences of households in TAZs

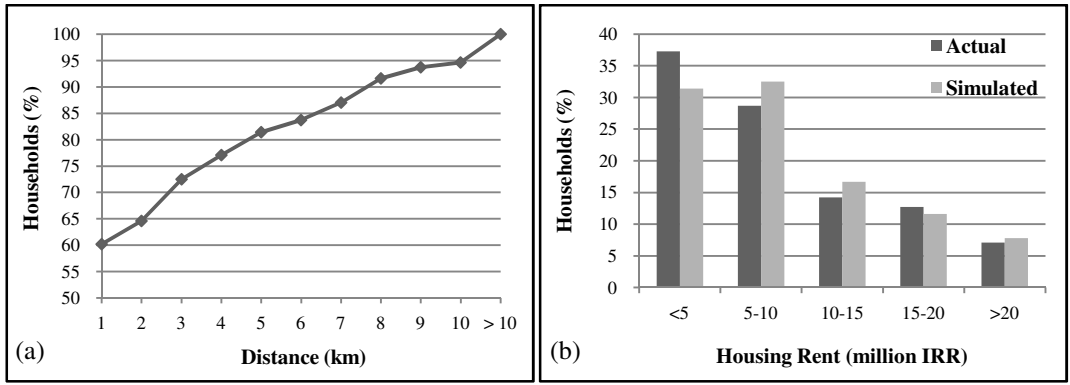

Figure (9): Distribution of households by: (a) distance between their simulated and actual residences; (b) housing rent of their actual and simulated residences

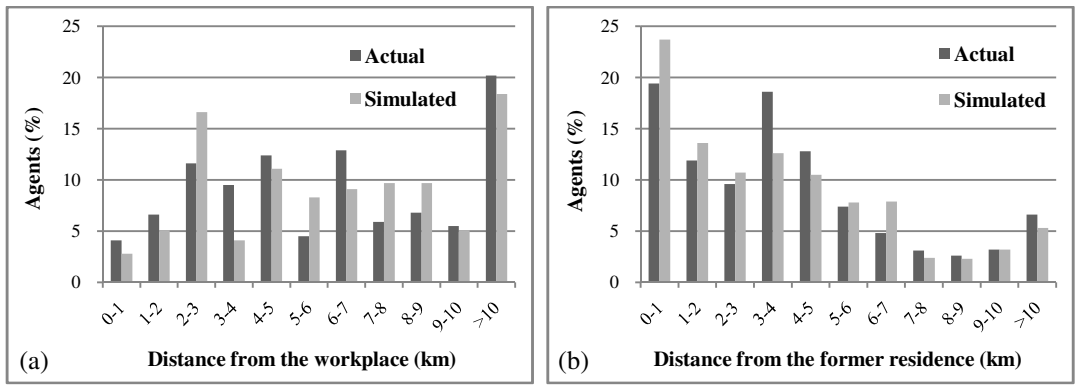

Figure (10): Distribution of households by: (a) distance from their workplace; (b) distance from their former residence

Table (5): Comparison of the actual and simulated residences of households

| | Households % |
|---|---|
| Simulated residence is identical to the actual residence | 59.3 |
| The actual residence is one of the simulated residential alternatives | 72.8 |
| Simulated residence is located in the adjacency of the actual residence | 67.5 |
| Simulated residence is located in distance of less than 5 km from the actual residence | 81.4 |
| Simulated residence is located in distance of more than 10 km from the actual residence | 5.4 |
| Rent of simulated residence is in the range of 85% to 115% of the rent of actual residence | 84.0 |
| Accessibility of simulated residence to the public facilities is in the range of 85% to 115% of accessibility of the actual residence | 78.1 |
| Accessibility of simulated residence to the highway network is in the range of 85% to 115% of accessibility of the actual residence | 84.5 |
| Accessibility of simulated residence to the public transit is in the range of 85% to 115% of accessibility of the actual residence | 70.6 |
| Air pollution level of simulated residence is identical to the actual residence | 79.2 |
| Noise pollution level of simulated residence is identical to the actual residence | 76.3 |

## 6. Discussion and Conclusion

In this paper, a novel two-step agent-based approach has been developed to simulate residential location choice of tenants. Tenants are considered as agents who compete with each other to reside in one of their preferred residential alternatives. They first select their desired residential alternatives according to their residential criteria and preferences using NSGA-II. In this step, they do not consider residential capacity of zones and residential alternatives of the other agents. Then, they select their final residential zone in a competition with other agents who look for the same zone at the same time period. Results show that the proposed approach has considerably simulated major residential patterns of different socio-economic categories of tenants in Tehran. For example, simulated residential choice behavior of agents with respect to accessibility to various public facilities and transportation services, air and noise pollutions and traffic restrictions are significantly consistent with the expectations. Also, validation results of the proposed approach with a sample of tenants shows that the approach has correctly simulated the actual residential zone of 59.3% of tenants. In addition, for more than 70% of sample households, various attributes of their actual residential zone including accessibility to public facilities and transportation services, air and noise pollutions and housing rent has simulated with errors of less than 15%. All of these results suggest the remarkable performance of the proposed approach.

The proposed approach has several advantages which lead to more closeness of the simulation of residential location choice of agents to the reality. First, due to using a multi-objective decision making method, NSGA-II, various conflicting objectives can be considered in the residence choice process of agents. There is no limitation on the number and type of objectives used in this method. The method also allows modeler to consider different objectives according to agents' preferences. In addition, the method can be used with a huge number of residential options. Second, In contrast with the conventional residential location choice models, agents do not essentially select the best residential option (the global optimum). They select a finite number of their desired residential alternatives with respect to their criteria and preferences and finally select their residence among these alternatives by competition with other agents.

The proposed residential location choice approach can be used in different urban applications. It can be considered as a main component of the land use and transportation models. It remarkably helps urban planners to investigate the residence choice behavior of individual households and consequently the spatial distribution of different socio-economic categories of households. Therefore, effects of various urban policies on these concerns can be investigated. Also, the proposed approach can be used for determining tenants who are unable to reside in any zone and possibly have to move to the surrounding cities of Tehran. Moreover, socio-economic structure of the population resided in different zones can be revealed and used for different purposes such as the housing subsidies to specific groups of population.

Although the proposed approach has been implemented in Tehran as a case study, it is a general approach which can be adapted with the residential location choice process in other developing and developed countries. The proposed multi-objective decision making method, NSGA-II, allows modelers to define different preferred criteria and objectives in accordance with the residential location choice in their study area. Also, although the criteria including number of members, income level, and number of children are considered in the proposed residential competition mechanism in Tehran, other criteria such as ethnicity, professional status, age and education level of the head of household can be considered in other areas in conformity with their cultural and socio-economic conditions.

Various aspects of the proposed approach need further considerations and developments. In this paper, the approach has been used by relatively large spatial units, TAZs, which may leads to modifiable areal unit problem (MAUP). This means that changes in the size or configuration of zones can affect the results (Openshaw & Taylor, 1979 ; Martinez et al., 2009). Therefore, MAUP effects must be carefully evaluated by using smaller spatial units such as parcels, census blocks or larger units resulting from their aggregations. Also, in this study, Euclidian distance has been used for measuring the proximity to different opportunities. Due to use of relatively large spatial units, TAZs, in this study, the use of Euclidian distance seems to be appropriate. It seems that people perceive proximity as a straight line (Euclidian distance) at the large spatial levels such as regions and TAZs. Also, the Euclidean distance is strongly correlated with the network distance at the census tract level in metropolitan areas (Apparicio et al., 2008). It is recommended that the network distance be used in future studies, especially when smaller spatial units such as census blocks or parcels are used. This measure may more accurately correspond to human perceptions of accessibility to different opportunities at the parcel level, because it measures the road distance between the parcel and the opportunity. However, calculation of the network distance is computationally intensive and requires greater user effort and knowledge for data input and preparation (Sander et al., 2010). The use of network distance in metropolitan areas such as Tehran which is composed of very dense transport networks may significantly reduce the computational performance of the proposed approach. In addition, only residential location choice of tenant households is simulated in this research, but the approach can be developed to include all categories of households. Finally, modules for determination of housing rent in a bid-rent framework and simulation of housing supply can be added to the approach in the future.